\documentstyle[psfig,12pt]{article}
\def\TL{\hfil$\displaystyle{##}$}
\def\TR{$\displaystyle{{}##}$\hfil}

\def\TT{\hbox{##}}
\def\seqalign#1#2{\vcenter{\openup1\jot
  \halign{\strut #1\cr #2 \cr}}}

\def\comment#1{}
\def\fixit#1{}

\def\overleftrightarrow#1{\vbox{\ialign{##\crcr
     $\leftrightarrow$\crcr\noalign{\kern-0pt\nointerlineskip}
     $\hfil\displaystyle{#1}\hfil$\crcr}}}

\def\lsim{\mathrel{\mathstrut\smash{\ooalign{\raise2.5pt\hbox{$<$}\cr\lower2.5pt\hbox{$\sim$}}}}}
\def\gsim{\mathrel{\mathstrut\smash{\ooalign{\raise2.5pt\hbox{$>$}\cr\lower2.5pt\hbox{$\sim$}}}}}

\def\sqr#1#2{{\vcenter{\vbox{\hrule height.#2pt
         \hbox{\vrule width.#2pt height#1pt \kern#1pt
            \vrule width.#2pt}
         \hrule height.#2pt}}}}

\def\href#1#2{#2}  

\def\lbldef#1#2{\expandafter\gdef\csname #1\endcsname {#2}}
\def\eqn#1#2{\lbldef{#1}{(\ref{#1})}%
\begin{equation} #2 \label{#1} \end{equation}}
\def\eqalign#1{\vcenter{\openup1\jot
    \halign{\strut\span\TL & \span\TR\cr #1 \cr
   }}}

\def\comment#1{  \begin{raggedright}{\tt [#1]}\end{raggedright}}
\def\fixit#1{}
\font\cmss=cmss10 \font\cmsss=cmss10 at 7pt

\def\IB{\relax\hbox{$\inbar\kern-.3em{\rm B}$}}
\def\IC{\relax\hbox{$\inbar\kern-.3em{\rm C}$}}
\def\ID{\relax\hbox{$\inbar\kern-.3em{\rm D}$}}
\def\IE{\relax\hbox{$\inbar\kern-.3em{\rm E}$}}
\def\IF{\relax\hbox{$\inbar\kern-.3em{\rm F}$}}
\def\IG{\relax\hbox{$\inbar\kern-.3em{\rm G}$}}
\def\IGa{\relax\hbox{${\rm I}\kern-.18em\Gamma$}}
\def\IH{\relax{\rm I\kern-.18em H}}
\def\IK{\relax{\rm I\kern-.18em K}}
\def\IL{\relax{\rm I\kern-.18em L}}
\def\IP{\relax{\rm I\kern-.18em P}}
\def\IR{\relax{\rm I\kern-.18em R}}
\def\Z{\relax\ifmmode\mathchoice
{\hbox{\cmss Z\kern-.4em Z}}{\hbox{\cmss Z\kern-.4em Z}}
{\lower.9pt\hbox{\cmsss Z\kern-.4em Z}}
{\lower1.2pt\hbox{\cmsss Z\kern-.4em Z}}\else{\cmss Z\kern-.4em
Z}\fi}

\def\II{\relax{\rm I\kern-.18em I}}

\def\tilde{\widetilde}

\def\inbar{\,\vrule height1.5ex width.4pt depth0pt}

\def\a{\alpha}
\def\b{\beta}

\def\d{\delta}
\def\e{\epsilon}

\def\th{\theta}

\begin{document}
\baselineskip=15.5pt
\pagestyle{plain}
\setcounter{page}{1}

\begin{titlepage}

\begin{flushright}
PUPT-1949 \\
CALT-68-2296\\
CITUSC/00-050\\
hep-th/0009140
\end{flushright}
\vfil

\begin{center}
{\huge The Hagedorn transition in}
\vskip0.4cm
{\huge non-commutative open string theory}
\end{center}

\vfil
\begin{center}
{\large S. S. Gubser,$^{1}$ S. Gukov,$^{1,2,3}$ I. R. Klebanov,$^{1}$
M. Rangamani,$^{1,2,3}$}
\vskip0.1cm
{\large and E. Witten$^{2,3,4}$}
\end{center}

$$\seqalign{\span\TL & \span\TT}{
^1 & Joseph Henry Laboratories, Princeton University, Princeton,
NJ 08544
 \cr\noalign{\vskip1\jot}
^2 & Department of Physics, Caltech, Pasadena, CA 91125
 \cr\noalign{\vskip1\jot}
^3 & CIT-USC Center for Theoretical Physics, Los Angeles, CA
 \cr\noalign{\vskip1\jot}
^4 & School of Natural Sciences, Institute for Advanced Study
 \cr\noalign{\vskip-1\jot}
   & Olden Lane, Princeton, NJ 08540
 \cr\noalign{\vskip1\jot}
}$$
\vfil

\begin{center}
{\large Abstract}
\end{center}

\noindent
 The Hagedorn transition in non-commutative open string theory (NCOS)
is relatively simple because gravity decouples.  For NCOS theories in
no more than five spacetime dimensions, the Hagedorn transition is
second order, and the high temperature phase involves long, nearly
straight fundamental strings separating from the D-brane on which the
NCOS theory is defined. Above five spacetime dimensions interaction effects
become important below the Hagedorn temperature. Although this
complicates studies of the transition, we believe that the
high temperature phase again involves long strings liberated from
the bound state.

\vfil
\begin{flushleft}
September 2000
\end{flushleft}
\end{titlepage}
\newpage
\section{Introduction}
\label{Introduction}

Just as non-commutative field theories ({\it i.e.}{} quantum field
theories on non-commutative spaces) can be obtained as certain limits
of D-branes with background magnetic field \cite{SW}, non-commutative open
string (NCOS) theories are defined as a special limit of
Type II D-branes with a uniform electric field \cite{SST,GMMS}.
The bosonic part of the world volume action has the standard form
\eqn{wsact}{
S= \int_\Sigma d^2 \sigma \left [
{1\over 4\pi\alpha'} \left (
(\partial_a X^0)^2 - (\partial_a X^1)^2\right ) -
{1\over 4\pi\alpha'_t}\sum_{i=2}^9 (\partial_a X^i)^2
\right ] + E \oint_{\partial \Sigma} X^0 {\partial\over \partial \sigma} X^1
\,.
}
 In the NCOS limit the electric field
approaches its critical value, $E \to E_c = 1/ 2 \pi \a'$, and $\a' \to
0$ in such a way that the effective
tension of an open string stretched along
the direction of the electric field remains finite:
  \eqn{apDef}{
   \alpha'_{\rm eff} = \a' {E_c^2 \over E_c^2 - E^2} \,.
  }
There is a similar rescaling of the interaction strength \cite{GKP}:
\eqn{inter}{
G_o^2 = g_{\rm str} \sqrt{E_c^2 - E^2\over E_c^2}\ ,
}
and the open
string coupling $G_o$ is held fixed in the NCOS limit.
The inverse-tension parameter for the transverse directions, $\alpha'_t$,
is finite from the start and independent of $E$ (it is convenient to
set $\alpha'_t = \alpha'_{\rm eff}$).
A remarkable property of this
limit is that, even though $g_{\rm str}\to\infty$, the
closed strings decouple from the open strings \cite{SST,GMMS}. Therefore,
the NCOS is a non-gravitational string theory.

This definition of NCOS theories leads naturally to a space-time
where the space and time directions don't commute,
{\it i.e.}{}, $[X^0, X^i] = i \theta^{0i}$. Unlike the situation in
non-commutative Yang-Mills theories, here the
non-commutativity scale, $\mid \th \mid$, is intrinsically tied to the
string scale. This implies that in order to make sense of the notion of a
non-commuting space/time manifold, we would have to first give precise
meaning to the notion of Einsteinian spacetimes down at the string scale.

The relation \inter\ implies that $g_{\rm str}\to \infty$ in the NCOS
limit. Therefore, S-duality may be used to map NCOS to D-brane systems
at weak string coupling \cite{GMSS,KM}. A particularly simple example
of such duality is $1+1$-dimensional NCOS, which is found to be dual to
maximally supersymmetric $U(N)$ gauge theory with one unit of electric
flux. The open string coupling is $G_o^2 = 1/N$; it becomes weak in
the large $N$ limit. Therefore, the $1+1$-dimensional NCOS provides a
new example of duality between large $N$ gauge theory and strings. The
fact that we find open strings rather than closed is related to the
presence of the electric flux tube which binds the $N$ D-strings.
Massive open strings are dual to the excitations of this theory where
locally $SU(N)$ is broken to $SU(N-1) \times U(1)$. The massless open
strings are dual to the $U(1)$ part of the spectrum (the overall
vibrations of the bound state), and the duality with the gauge theory
predicts that the massless states decouple. In \cite{KM} this
prediction was confirmed by explicit NCOS calculations.  A further
check on the duality performed in \cite{KM} involves the high-energy
behavior of the massive amplitudes: it is found to exhibit the same
power-law fall-off as expected from the gauge theory.

Another classic way of subjecting strings to extreme conditions is to
heat them up to a high temperature.  For conventional superstring
theory this was extensively studied in the late 80's
\cite{Kogan,Sathiapalan,AW} and afterwards (see for instance
\cite{Bowick,Tan,Lowe}).  A complicating factor in these papers is that it
is difficult to study thermodynamics of gravitating
systems. Nevertheless, a coherent picture has emerged suggesting a
first-order phase transition happening well below the Hagedorn
temperature \cite{AW}.

In this paper we study the thermodynamics of NCOS in various dimensions.
Just like any other superstring theory, NCOS theory exhibits
a Hagedorn density of states:
  \eqn{RhoBehave}{
   \rho (m) \sim m^{-9/2} e^{m \over T_H} \,,
  }
with the scale for the Hagedorn temperature set by $\alpha'_{\rm eff}$:
  \eqn{thagedorn}{
   T_H = {1 \over \sqrt{8 \pi^2 \alpha'_{\rm eff}}}
  \ .}
In particular, the partition function of
NCOS theory appears to diverge above the temperature $T=T_H$,
where the Hagedorn
transition is expected to take place.  Our goal is to understand the
physics of this transition and describe the thermodynamics of NCOS
theory at $T > T_H$.
Since the NCOS theories are decoupled from gravity, we will
not face the usual difficulties associated with gravitational
thermodynamics. Furthermore, at least in 1+1 dimensions the dual
gauge theory provides an important guide to what happens at the transition.
Here we find that the transition is to a phase where some
finite fraction of the strings are freed from the bound
state, {\it i.e.}{} where the theory
enters the Higgs branch $SU(N)\to SU(N-K)\times U(1)^K$.
A calculation of the free energy below and above the transition shows
that it is second order.

Guided by the intuition from the $1+1$-dimensional case we proceed to
$p+1$ dimensions. For $p>1$, S-duality works differently, but
on the NCOS side we may still think of some density of F-strings
bound to a D$p$-brane. We will show that for all $p<5$
the Hagedorn transition is again second order
and is associated with liberation of
strings from the bound state.  For
$p \geq 5$ the entropy of non-interacting open strings
(and also the string length) diverges as $T$
approaches $T_H$ from below \cite{Eli}. 
This suggests that string interaction effects
become important already below $T_H$. Nevertheless, we will
argue that the high temperature phase again contains a finite
fraction of long strings liberated from the bound state.
In all these cases the theory slightly
above the transition appears to be effectively $1+1$-dimensional, with
a preferred direction chosen by the electric field.

In previous work \cite{GMMS} it was suggested that there is a change
in the behavior of zero-temperature
NCOS at $p=7$, where non-planar amplitudes begin to diverge at
$k^2 =0$ ($k$ is the momentum flowing in the closed string channel). 
We calculate a cross-section
for graviton production and confirm that NCOS theories do not
decouple from gravity for $p\geq 7$.
Our work further shows that, at finite temperature, 
there is new physics appearing in the NCOS theory at
a lower dimension, $p=5$: interaction effects
become important already below $T_H$. A special role of $p=5$ in open string
thermodynamics was noted earlier in \cite{Eli}.

Other authors have recently studied phases of NCOS theories and
Hagedorn behavior of string theories decoupled from gravity
\cite{HarmarkZero,HarmarkOne,HarmarkTwo,Sahakian}.  These works
focused primarily on results derivable from supergravity.  The current
work takes the rather different approach of examining the free energy
of bound states in a field theory approximation.  The relevant
temperatures for our analysis are so low that the near-extremal
supergravity solutions are highly curved on the string scale and hence
unreliable.

There is also an extensive literature on Hagedorn behavior in
asymptotically free gauge theories.  For important early contributions
to the subject, see \cite{PolQCD,Susskind,thorn}.  The current work
focuses on perturbative string techniques rather than field theory.
However, some information about strongly coupled gauge theories may be
extracted from our results, particularly in the 1+1-dimensional case.

\section{NCOS thermodynamics for $T<T_H$}
\label{LowT}

In order to obtain a reliable picture of the thermodynamics of NCOS
theory for $T<T_H$ directly from the free string spectrum, two
conditions must pertain.  First, the open strings should interact
weakly with one another.  Second, the cubic coupling
$\langle\phi\phi\sigma\rangle$ between an incipient thermal tachyon
$\phi$ and the radius $\sigma$ of the Euclidean time direction,
which played a crucial role in the analysis of \cite{AW}, is
absent. This is because $\sigma$ represents a closed string
(gravitational mode), which decouples according to the arguments of
\cite{SST,GMMS}.  This is the essential difference between
the Hagedorn transition for NCOS theory and for critical string theory
\cite{AW}.  Whereas essentially gravitational effects drive the
Hagedorn transition first order in critical string theory, we will see
that in NCOS theory the transition remains second order.

The free string analysis proceeds in a similar way regardless of the
spatial dimension $p$ in which the open strings live.
The calculation of free energy of non-interacting open strings on
a D$p$-brane, which is not affected by the non-commutativity,
was carried out in \cite{Eli}. We will largely rederive
their results and adapt them for our purposes. The principal
result is that
the free energy is analytic in $T$ for $T<T_H$, and that the leading
non-analytic behavior in the expansion of $F$ around $T=T_H$ is
  \eqn{FBehave}{
   F \sim \left\{ \seqalign{\span\TR\quad & \span\TT}{
    \hbox{(analytic in $t$)} + t^{{7-p \over 2}} + \ldots
       & for $p$ even  \cr
    \hbox{(analytic in $t$)} + t^{{7-p \over 2}} \log t + \ldots
       & for $p$ odd} \right.
  }
 where $t = (T_H-T)/T_H$.  There are two equivalent means of obtaining
this result.  First, one may directly evaluate the annulus diagram in
the Matsubara formalism where the Euclidean time direction is compact with
circumference $\b = T^{-1}$.
The one loop free energy for a D$p$-brane with an electric
field of strength $E$ turned on is \cite{Tseytlin}
  \eqn{zncos}{\eqalign{
   Z_{\rm single\ string} &= -c_1 \int_0^{\infty} {dt \over t^{p+3 \over 2}}
    \vartheta_2 \Big( 0 \vert {i \b^2 \over 2 \pi^2 \alpha'_{\rm eff} t} \Big)
    \Big[ {\vartheta_2 ( 0 \vert it) \over \vartheta_1' ( 0 \vert it)} \Big]^4
     \cr
     &= -{c_1 \over (2 \pi)^4}
    \int_0^{\infty} {d \tau \over \tau^{9 - p \over 2}}
    \vartheta_2 \Big( 0 \vert {i \b^2 \tau \over 2 \pi^2 \alpha'_{\rm eff} }
      \Big)
    \Big[ {\vartheta_2 ( 0 \vert i \tau) \over \vartheta_1' ( 0 \vert i\tau)}
     \Big]^4 \,.}
  }
 In the first line we have used $t$ as the modular parameter of the
cylinder.  In the second line we substitute $\tau = 1/t$, which is the
usual closed string modular parameter.  The expressions in \zncos\ are
exact even away from the NCOS limit, provided we use the definition
\apDef\ and neglect coupling to closed strings.  They are identical
to the partition function of ordinary open superstrings on a D$p$-brane,
only with $\alpha'$ replaced by $\alpha'_{\rm eff}$.  The constant
$c_1$ is given as ${ V \b \pi^4 \over 2 (2 \pi)^5 (2 \pi \alpha'_{\rm
eff})^5}$.  The reduced Hagedorn temperature for the partition
function was noted in \cite{Fradkin}.

The non-analytic behavior arises from a divergence in the modular
integral at large~$\tau$ (the long cylinder limit).  Near
$T_H$, we can use the large $\tau$ asymptotics of the
$\vartheta$-functions to obtain
  \eqn{fncos}{
   F_{NCOS} (T \approx T_H) \sim
    -\int^\infty {d \tau \over \tau^{9-p \over 2}} e^{({1 \over T_H^2} -
     {1 \over T^2}) \tau} \,,
  }
 from which the claimed analyticity for $T<T_H$ and the leading
non-analyticity quoted in \FBehave\ are evident.

An equivalent, ``elementary'' approach to obtain the same result is to
plug \RhoBehave\ into the standard formula for a partition function:
  \eqn{znAgain}{\eqalign{
   Z_{\rm single\ string} &= \sum_{\rm states} e^{-E/T}  \cr
    &= \sum_{i \in {\cal H}^\perp_o} \int {d^p k \over (2\pi)^p}
         e^{-\sqrt{k^2+m_i^2}/T}  \cr
    &\sim \int_0^\infty dm \, \rho(m) (mT)^{p/2} e^{-m/T}  \cr
    &\sim \int^\infty dm \, m^{(p-9)/2}
       \exp \left[ m \left( {1 \over T_H} - {1 \over T} \right) \right] \,,
  }}
 where in the third step we have made an approximation to the momentum
integration which becomes exact in the limit of 
large masses $m_i$ \cite{Eli}.
Evaluating the last integral leads again to \FBehave.

Note that the free energy is finite at $T=T_H$ for $p<7$ and diverges
logarithmically for $p=7$.  The entropy, $S = -\partial F/\partial T$,
remains finite only for $p<5$, and diverges logarithmically for $p=5$.
For a single long string, the entropy is proportional to the length of
the string.  Hence when $t = (T_H-T)/T_H$ is small, the total entropy
is proportional to the r.m.s.\ length of the excited open strings,
$l_{NCOS}$, times the average number of these strings per unit volume,
$\rho_{NCOS}$.\footnote{It has been argued (see \cite{Bowick,Tan,Lowe} and
references therein) that near the Hagedorn temperature, strings tend
to merge, so that the average number of strings per unit volume
decreases while the average length increases.  For our purposes, only
the product $\rho_{NCOS} l_{NCOS}$ is relevant.}  As $T \to T_H$ from
below, we have the scalings
  \eqn{length}{
   \rho_{NCOS} l_{NCOS} \sim \left\{ \eqalign{\hbox{(finite)}
    \quad\hbox{for $p<5$}  \cr
    -\log t \quad\hbox{for $p=5$}  \cr
    1/\sqrt{t} \quad\hbox{for $p=6$}} \right.
  }
 and so on.

The quantity $\rho_{NCOS} l_{NCOS}$ is the average density of string
at any given point.  As long as this quantity remains finite, the
effects of interactions may be suppressed by taking $G_o$ sufficiently
small.  Thus for $p < 5$ one can ensure that string
interactions are never significant, but for $p \geq 5$ they eventually
will be.  A figure of merit to measure the strength of string
interactions is $\eta = G_o \rho_{NCOS} l_{NCOS}$.  We work in units
where $\alpha'_{\rm eff} = 1$ to make $\eta$ dimensionless.  The free
energy for $T<T_H$, neglecting interactions, is order $G_o^0$.  
Interactions make a
contribution of order $\eta^2$ to the free energy.  Thus interactions
become important when $\eta \gsim 1$, which is to say $\rho_{NCOS}
l_{NCOS} \gsim 1/G_o$.  One can now use \length\ to make a rough
estimate of the temperature at which string interactions matter.  For
$p=5$ this temperature is $t \sim e^{-{\rm const}/G_o}$, while for $p=6$ it
is $t \sim G_o^2$.  

In the next sections, we will propose that the physics above $T_H$
involves gradual emission of long strings.  We will assume that the
free string picture is valid up to $T=T_H$: thus the discussion seems to be
limited to $p<5$.  Note however that for $p=5,6$, the entropy of the
open string gas at the temperature where interactions become important
is of order $1/G_o$.  For weak string coupling, this is still much
smaller than the entropy in the liberated string phase which, as we
show in section~\ref{Higher}, is of order $1/G_o^2$.  So we speculate
that long string liberation starts taking place near $T_H$ for $p=5,6$
as well.

The calculations that we present for $p < 5$ are clean because we can
work in a limit where free string theory applies.  The string
liberation transition may still occur away from zero coupling,
although it is possible that the transition becomes first order.  The
additional complication for $p=5,6$ is that there is no limit in which
free string theory applies uniformly.

\section{Two-Dimensional NCOS theory at $T>T_H$}
\label{HighT}

In this section we focus on the specific example of the
two-dimensional NCOS theory, $p=1$.  In this case one can use Type IIB
S-duality to describe a $D1$-brane with a near-critical electric field
as a $(1,N)$ bound state \cite{SchwarzStrings,WittenBound} where the
number, $N$, of $D1$-branes is related to the open string coupling
constant, $G_o^2 = 1/N$.  At low energies this system behaves as
$SU(N)$ two-dimensional super-Yang-Mills theory with one unit of
electric flux and coupling constant
  \eqn{gym}{
   g_{YM}^2 = {N^2 \over \alpha'_{\rm eff}} \,.
  }
 In this dual picture non-commutative open strings can be identified
with excitations corresponding to the Higgsing $SU(N) \to SU(N-1)
\times U(1)$.  Indeed, to create an island (of size $L$) of the Higgs
phase costs an energy \cite{GMSS,KM}
  \eqn{tens}{
   E = L \Big( {g_{YM}^2 \over 4 \pi (N-1)}
    - {g_{YM}^2 \over 4 \pi N} \Big)
    \approx {L g_{YM}^2 \over 4 \pi N^2}
    = {L \over 4 \pi \alpha'_{\rm eff}} \,.
  }
 In the last equality we used the relation \gym\ between Yang-Mills
coupling constant and the tension of open strings.  In terms of the
$(1,N)$ bound state, \tens\ represents the energy to have a
D-string split off from the bound state and run parallel to
it for a distance $L$ before rejoining.

At finite temperature, there is a gain in entropy when a string splits
off from the bound state, due to small fluctuations of the string.
Since a long string in light-cone gauge is described by a free
supermultiplet, this entropy is $S = 4\pi L T$ (in the dual gauge theory
it comes from the $U(1)$ part of the Higgsed gauge group 
$SU(N-1) \times U(1)$).
The corresponding free energy of these light modes is $F=-2\pi L T^2$.
Therefore, the total free energy of a string liberated from the bound state,
  \eqn{FreeF}{
   F_{\rm liberated\ string} =
     L \left( {1 \over 4 \pi \alpha'_{\rm eff}} - 2 \pi T^2 \right) \,,
  }
 vanishes precisely at the Hagedorn temperature:
  \eqn{THAgain}{
   T_H = {1 \over \sqrt{8 \pi^2 \alpha'_{\rm eff}}} =
    {1 \over \sqrt{8} \pi} {g_{YM} \over N} \,.
  }

We would like to interpret the Hagedorn transition as the liberation
of fundamental strings parallel to the electric field from the $(N,1)$ bound state.
This interpretation is satisfying in that the Hagedorn transition is
generally associated with the temperature at which it is favorable to
create long closed strings (see for instance \cite{Bowick,Tan}).  If we
compactify in the direction of the electric field, then the liberated
strings are precisely those long closed strings.  What is special
about the NCOS limit is that in the near-critical electric field, the
closed string are allowed to wind only in one direction.  Furthermore,
since their tension away from the bound state is $\alpha'$, they are
nearly straight: the massless $U(1)$ degrees of freedom represent only
slight fluctuations.

A crucial aspect of the analysis is that, once one string has been
liberated, the Hagedorn temperature of the NCOS theory on the
$(N-1,1)$ bound state is slightly higher: after freeing one string, we
have
  \eqn{THNew}{\eqalign{
   \alpha'{}_{\rm eff}^{\rm new} &= {(N-1)^2 \over g_{YM}^2}  \cr
   T_H^{\rm new} &= {g_{YM} \over \sqrt{8} \pi} {1 \over N-1} \,.
  }}
 In order to free another fundamental string, we must increase $T$ to
$T_H^{\rm new}$.  The analysis in \tens\ and \FreeF\ carries over
without change to this case, and the Hagedorn temperature of the bound
state increases again.  Thus we have good control over the physics
above the original $T_H$: the liberated fundamental strings are only
slightly fluctuating, and the NCOS strings attached to the bound state
remain at or below their Hagedorn transition.

It is possible to summarize the analysis in a way that will generalize
easily to other cases.  Suppose $k$ out of the $N$ fundamental strings
have been liberated, $k \gg 1$.  The free energy per unit length of
the total system, consisting of the $(N-k,1)$ bound state plus the $k$
liberated strings, is
  \eqn{FkForm}{\eqalign{
   {F_k \over L} &= {1 \over 2\pi\alpha'} \sqrt{(N-k)^2 +
     {1 \over g_{\rm str}^2}} + {k \over 2\pi\alpha'} -
     2\pi k T^2 + O(1)  \cr
    &\approx {N-k \over 2\pi\alpha'} \left( 1 +
     {1 \over 2 g_{\rm str}^2 (N-k)^2} \right) +
     {k \over 2\pi\alpha'} - 2\pi k T^2  \cr
    &= {N \over 2\pi\alpha'} - 2\pi N T^2 +
     {1 \over 4\pi \alpha' g_{\rm str}^2 (N-k)} +
     2\pi (N-k) T^2 \cr
    &\geq {N \over 2\pi\alpha'} - 2\pi N T^2 +
     T \sqrt{2 \over \alpha' g_{\rm str}^2} \,.
  }}
 In the first line of \FkForm, we have summed up the total tension of
the $(N-k,1)$ bound state, the total tension of the $k$ liberated
strings, the free energy of the fluctuations of those $k$ strings, and
the $O(1)$ free energy coming from fluctuating open strings attached
to the bound state.\footnote{The total tension of each liberated
string is indeed $1/2\pi\alpha'$.  The tension $1/4\pi\alpha'_{\rm
eff}$ commonly ascribed to these strings is their {\it net} tension,
over and above the tension they would have added to the bound state
had they remained bound.}  The symbol $O(1)$ means, more precisely,
that this contribution to the free energy is a finite quantity of
order $1$ times $LT^2$.  In the second line of \FkForm\ we have
expanded the square root for $g_{\rm str} (N-k) \gg
1$, and in the last line we have used the arithmetic-geometric mean
inequality.  Equality holds in the last line iff
  \eqn{NkDiff}{
   N-k = {1 \over \sqrt{8\pi^2 \alpha'} g_{\rm str} T} \,.
  }
Transforming to rescaled NCOS variables, we find that
the fraction of liberated strings is
  \eqn{NuDef}{
   \nu = {k \over N} = {T-T_H \over T} \ ,
  }
where $T_H$ is the original Hagedorn temperature
defined in \THAgain.

 From \FkForm\ we immediately read off the total free energy for
$T>T_H$:
  \eqn{TotalF}{
   {F \over L} = \inf_k {F_k \over L} =
    -2\pi N (T-T_H)^2 + O(1) \, .
  }
Another way to arrive at this formula is to note that
the entropy of the $k$ liberated strings is
\eqn{libent}{
S = -{d F\over dT} = 4 \pi k LT = 4 \pi N L (T-T_H)
\ .
}
Integrating this equation with the boundary condition that
$F/L$ is of order 1 at $T=T_H$ reproduces the result
\TotalF.

 Actually, since $k$ is a discrete variable, $F/L = -2\pi N (T-T_H)^2$
and $\nu = (T-T_H)/T$ only represent an approximation to a series of discrete
transitions, from $F_0$ to $F_1$ to $F_2$ and so on.  However, since we
are operating at large $N$, the discrete transitions are very closely
spaced, and can effectively be regarded as a single continuous
transition.
At some level, the approach we have taken is
only meaningful in the large $N$ limit: we have examined
various competing minima of the free energy, corresponding to
different numbers of liberated fundamental strings, at a completely
classical level, ignoring the fact that in one spatial dimension
strong infrared fluctuations smooth out any non-analyticity in the
free energy.  What saves the day is large $N$: when a finite fraction
of fundamental strings have been liberated, their fluctuations
``average out'' to an extent such that $\nu$ is a good order parameter
for the transition.  One should not take too seriously the literal
picture of a single string peeling off the bound state at $T=T_H$,
followed shortly thereafter by another, and then another; rather, the
free energy starts as an $O(1)$ quantity for $T<T_H$ and rises to
$O(N)$ through a transition in which fundamental strings are
collectively liberated.  There is {\it not}, after all, a series of
closely spaced first-order transitions---this would be in violation of
the general analyticity properties of the free energy in one spatial
dimension---instead, the maximally smoothed free energy has the form
\TotalF, which indicates a discontinuity $\Delta C = 2\pi N$ at
$T=T_H$.  This is essentially the classic picture of a second order phase
transition, only with integer critical exponents that just barely
avoid the typical singularity in the specific heat.  

Note that, since the equilibrium condition reads
\eqn{alwaysH}{
{g_{YM}^2 \over 4\pi (N-k)^2 } - 2\pi T^2 =0\ ,
}
the open strings on the $(N-k,1)$ bound state are always at their
effective Hagedorn temperature (which depends on $k$ provided that
$g_{YM}$ is held fixed). Therefore, their contribution to $F/L$
is $a T^2$, where $a$ is a constant of order 1 which may be found by
evaluating \zncos\ directly at the Hagedorn temperature.
It is quite possible
that additional $O(1)$ contributions to $F$ arise when one considers
interactions of the liberated long strings.
Also, the interactions
could change the critical exponents by terms of order
$G_o^2=1/N$.  In principle, the
effects of interactions can be studied starting from the maximally
supersymmetric Yang-Mills description of the bound state.  The
liberated strings admit a matrix string description 
\cite{Motl,Banks,DVV}, while
the bound state represents a confining non-abelian sector of the
theory.  As far as we can tell, the total problem is quite formidable,
but some progress might be made via a lattice or DLCQ approach.

It is clear that as we increase the temperature, one unit of electric
flux in the dual super-Yang-Mills theory becomes unimportant.  Already
when the temperature is a finite multiple of $T_H \sim g_{YM}/N$ (say
$T = 2 T_H$), the free energy is dominated by the matrix string phase
(recall that we are mainly interested in the large $N$ limit).  When
$T \sim g_{YM}/\sqrt{N}$, the proper description of the system is no
longer matrix string theory plus a D1-f1 bound state, but rather a
single near-extremal black string solution in type IIB supergravity
\cite{KS,IMSY}.  The considerations of \cite{KS,IMSY} were applied
only to multiple, identical, (nearly) coincident branes, but their
conclusions should carry over to the current circumstance, because at
$T = g_{YM}/\sqrt{N}$, nearly all the D1-branes are in the matrix
string phase: $\nu = 1 - O(N^{-1/2})$.  The supergravity regime, then,
is described by \cite{KS,IMSY}
  \eqn{KSF}{
   F \sim L N^{3/2} {T^3 \over g_{YM}} \quad\hbox{for
    ${g_{YM} \over \sqrt{N}} < T < g_{YM} \sqrt{N}$.}
  }
 Finally, as we reach the 't~Hooft scale $T \approx g_{YM} \sqrt{N}$,
we end up with a gas of free photons, $N^2$ in number.  This crossover
is in the general class of correspondence points studied by Horowitz
and Polchinski \cite{hp}.  At very high temperature the free energy
looks like this:
  \eqn{HighTF}{
   F \sim L N^2 T^2 \quad\hbox{for $g_{YM} \sqrt{N} < T$.}
  }
 Historically, the Hagedorn transition was originally expected to be
essentially a deconfinement transition.  In the NCOS context, we see
that there are actually two other phases, or regimes, in between the
Hagedorn transition and the free gluon phase.  In a superficial
matching analysis, the transitions between the matrix string regime,
the supergravity regime, and the free gluon regime appear to be first
order.  It could easily be, however, that there is only a second order
transition, or no sharp transition at all, between these phases.  As
yet, we know of no method of analysis powerful enough to distinguish
among the possibilities.  For the transition into the supergravity
regime from below, one may hope that the perturbation of the matrix
string CFT by the DVV twist operator \cite{DVV} provides some hint of the
formation of a horizon.

\begin{figure}[p]
   \centerline{\psfig{figure=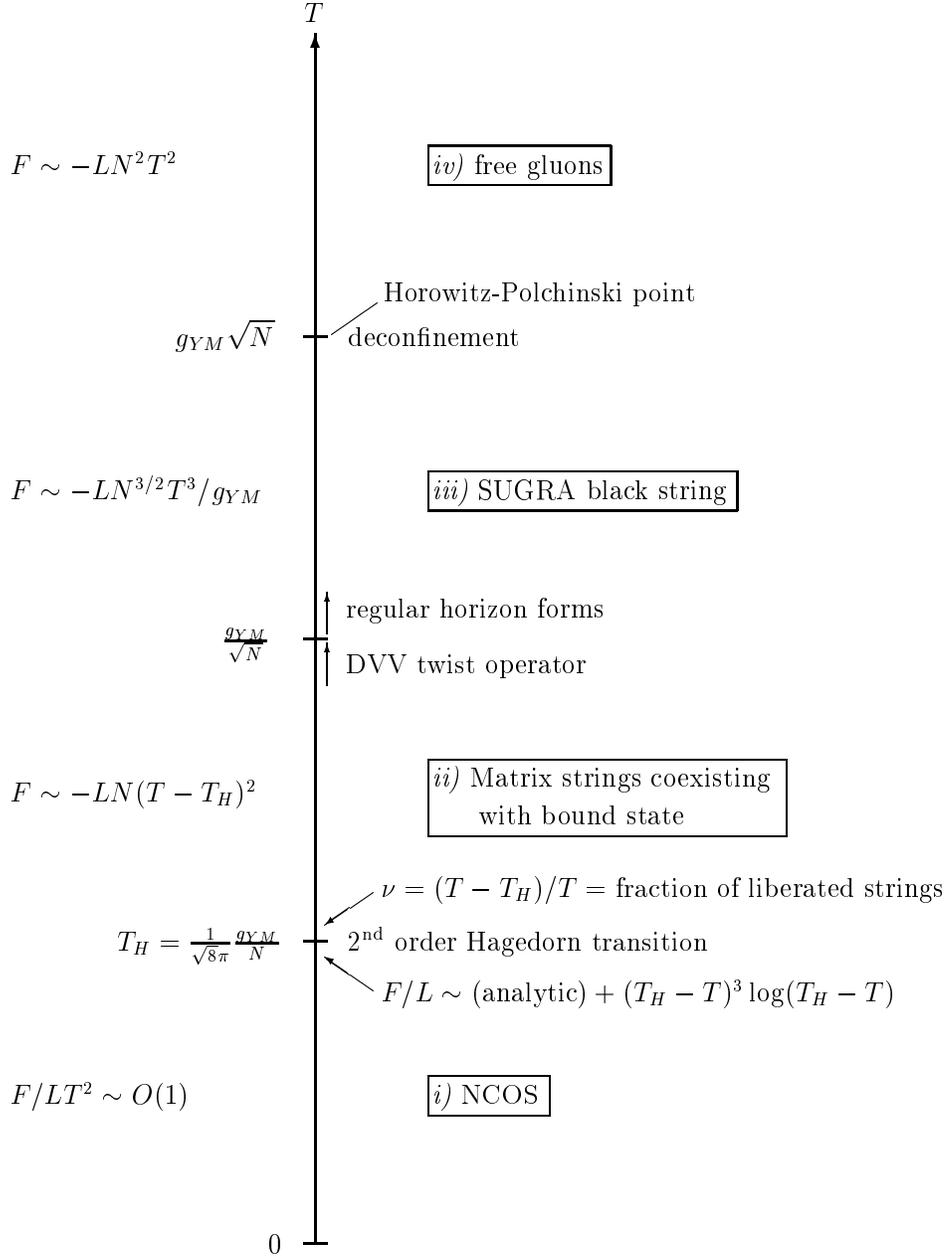,width=4.9in}}
   \caption{The four phases of $1+1$-dimensional NCOS theory.}\label{figA}
  \end{figure}

In summary, we find four different phases of $1+1$-dimensional
NCOS theory.  They are illustrated in figure~\ref{figA}.
 In more detail, we have

$i)$ In the NCOS phase, $F/LT^2$ is an analytic function of 
order $1$ (that
is, no factors of $N$).  String interactions are suppressed by $G_o^2
= 1/N$.  Without an understanding of the gauge theory and the
possibility of going to a Higgsed phase $SU(N-k) \times U(1)^k$, this
is the only part of the phase diagram we would be able to understand.

$ii)$ Above $T=T_H$, we gradually liberate more and more fundamental
strings from the bound state, so that very soon the system becomes
dominated by the matrix string phase. The open strings on the bound state
stay at their effective Hagedorn temperature: this temperature
adjusts as more strings are liberated.  The continuous transition so
described is the essential new physics of this paper.

$iii)$ At $T \approx {g_{YM} \over \sqrt{N}}$ significant departures
from conformal invariance and non-trivial interactions drive us into
the black string regime, where the thermodynamics is read off from a
regular horizon.

$iv)$ Above $T \approx g_{YM} \sqrt{N}$, the $N^2$ non-abelian gluons
(light D1-branes stretched between fundamental strings) are deconfined.

It is straightforward to extend our discussion to $(N,M)$
bound states corresponding to $SU(N)$ theory with arbitrary
number, $M$, of flux units.
In that case, open string coupling constant is given by:
$$
G_o^2 = {M \over N}
$$
and effective open string tension reads:
$$
{1 \over \alpha'_{\rm eff}} = g_{YM}^2 {M^2 \over N^2}
$$
The phase diagram of this system is similar to that of $(N,1)$
bound state: in particular, the same four phases appear.
The only difference is that the phase transitions between phases
$i)$ -- $ii)$ -- $iii)$ occur at different temperatures,
greater by a factor of $M$.
For instance, the Hagedorn temperature of such a theory is given by:
$$
T_H = {g_{YM} M \over \sqrt{8 \pi^2} N}
\ .$$

\section{Higher dimensional examples}
\label{Others}

Let us now elaborate on extensions of the ideas of the previous
section to systems in which the strings are allowed to move in more
than one spatial dimension (section~\ref{Higher}), or where not
strings but D$p$-branes become light (section~\ref{OM}).

\subsection{Thermodynamics of NCOS theories in higher dimensions}
\label{Higher}

The Hagedorn transition in higher dimensional NCOS theories
(up to $4+1$ dimensions) may be
understood in a manner similar to the situation in 1+1 dimensions.
The claim is that finitely above $T_H$, a finite fraction of
long strings are liberated from the bound state. This process is
gradual as in the previous case, and the fraction of long strings that
decouple may be computed in a similar fashion.
To see this we need the formula for the number of fundamental strings
per unit transverse volume bound
to a D$p$-brane with electric field $E$:
\eqn{nerel}{ {N \over V_t}  \sim (\alpha'_t)^{(1-p)/2}
{E \over g_{\rm str}\sqrt{ E_c^2 - E^2}}.}
One way to get this formula is to consider the D$p$-brane to be compactified on
a circle of radius $L$
in the direction of the electric field. Then the momentum conjugate
to the gauge field is quantized: $P_1=NL$. Equating this to the momentum
calculated from the Born-Infeld action, as in \cite{GKP}, gives \nerel.

Note that \nerel\ implies
$$ G_o^2 \sim {V_t (\alpha'_t)^{(1-p)/2}\over N}
\ .
$$
To fix the precise factor in this expression, consider the BPS
formula for the mass of the bound state of
$N$ fundamental strings and a D$p$-brane
wrapped over a transverse torus of volume $V_t$:
\eqn{BPSf}{ {L\over 2\pi \alpha'}
\sqrt{ N^2 + {V_t^2\over g_{\rm str}^2 (2\pi)^{2p-2}
(\alpha'_t)^{p-1} }} = {L\over 2\pi \alpha'} N
\left (1+ {V_t^2\over 2 N^2 g_{\rm str}^2 (2\pi)^{2p-2}
(\alpha'_t)^{p-1} } + \ldots \right )\ .
}
Calculating the energy required to free one fundamental string,  we get
\eqn{wounden}{ { L V_t^2\over 4\pi \alpha' g_{\rm str}^2 (2\pi)^{2p-2}
(\alpha'_t)^{p-1} } \left ({1\over N-1} - {1\over N}\right )
\rightarrow { L V_t^2\over 4\pi \alpha'_{\rm eff} G_o^4 N^2 (2\pi)^{2p-2}
(\alpha'_t)^{p-1} }
\ ,
}
where we have used
\eqn{scalrel}{ \alpha' g_{\rm str}^2 = \alpha'_{\rm eff} G_o^4\ .
}
\wounden\ should be equated to the energy of a closed string wound around the
direction of the electric field,
which is $L/(4\pi \alpha'_{\rm eff})$ \cite{KM}.
Thus, we find
\eqn{morep}{
G_o^2 ={V_t (\alpha'_t)^{(1-p)/2}\over (2\pi)^{p-1} N}
\ .
}
This formula shows that for $p>1$, $G_o^{-2}$ is not quantized, while
for $p=1$ it is.


Suppose we start with a D$p$-brane with a near-critical electric
field $E$, which loses a fraction of its long strings above $T_H$,
such that the resulting system is a D$p$-brane with a near-critical
electric field $E'$ and a bunch of free long strings.
Assuming that the resulting brane configuration is right at its
effective Hagedorn
temperature as before, we find the ratio
$$
{T \over T_H} = {\sqrt{E_c^2- E'^2} \over \sqrt{ E_c^2 - E^2}} \,.
$$
Using \nerel\ and the fact that both $E$ and $E'$ are near-critical,
we find the relation between temperature and the fraction of strings
remaining in the bound state:
\eqn{boundfrac}{
{N' \over N} = {T_H \over T}\ .
}
This universal result is in accord with what we found
in two-dimensional NCOS theory from its gauge theory dual,
{\it cf.} \NuDef.

Note that the transverse inverse-tension parameter, $\alpha'_t$, remains
fixed for $T>T_H$ because it does not depend on $E$.
Thus, we may simply  set $\alpha'_t = \alpha'_{\rm eff}$.
However, the effective parameter governing the $0$ and $1$ directions
starts decreasing as in the $1+1$-dimensional case. From \scalrel\ 
and \morep\ we find that
$$ \alpha'{}_{\rm eff}^{\rm new} =  \alpha'_{\rm eff}
\left( {N'\over N} \right )^2
\ .
$$
As in 1+1 dimensions, the condition for equilibrium of long strings at
temperature $T$ is
$$ {1\over 4\pi \alpha'{}_{\rm eff}^{\rm new}} - 2\pi T^2 =0
\ ,
$$
from which the relation \boundfrac\ follows.

One may be concerned that for $T>T_H$ there are two different effective
inverse-tension parameters: $\alpha'{}_{\rm eff}^{\rm new}$ for the
$01$ directions and $ \alpha'_{\rm eff}$ for the transverse directions.
Which one sets the effective Hagedorn temperature? The answer is that
it is $\alpha'{}_{\rm eff}^{\rm new}$, so that $T$ is the effective
Hagedorn temperature for $T>T_H$. The dispersion relation for open
strings is indeed asymmetric:
$$  (k_0^2- k_1^2)-
{\alpha'_{\rm eff}\over  \alpha'{}_{\rm eff}^{\rm new}}
\sum_{i=2}^p k_i^2 = {{\cal N}\over \alpha'{}_{\rm eff}^{\rm new}}
\ ,
$$
where ${\cal N}$ is the excitation level. We see that
$\alpha'{}_{\rm eff}^{\rm new}$ determines the mass spectrum.
Then following, for instance, the approach in \znAgain\ we find
that the effective Hagedorn temperature is
$(8\pi^2 \alpha'{}_{\rm eff}^{\rm new})^{-1/2}=T$. Therefore, the
free energy of the gas of open strings on the bound state is a finite
(for $p< 7$) quantity of order $1$, as far as the dependence on $G_o$
is concerned.

For $T>T_H$ the free energy is dominated by that of the
$N-N'$ free long strings:
$$ F= - 2\pi N L (T-T_H)^2\ .
$$
Using \morep\
we observe that this expression is extensive:
\eqn{exten}{ F= - L V_t (2\pi)^{2-p}
(\alpha'_{\rm eff})^{(1-p)/2} G_o^{-2} (T-T_H)^2
\ .
}
Just as for $p=1$, the free energy is
of order $G_o^{-2}$ for $T> T_H$.

Now we are in a position to complete the phase diagram of the higher
dimensional NCOS theories.
At very low temperatures, one has the
open string phase of the NCOS. This description breaks down at the
Hagedorn temperature \thagedorn, where one has a phase transition
beyond which
the temperature dependence is effectively two-dimensional although
the free energy remains extensive.
The details of the phase transition are dimension dependent.
For all $p<5$ the Hagedorn transition is second order.
In $p\geq 3$ the specific heat diverges as
$T\to T_H$ on the open string side \cite{Eli}. 

The 
free energy of non-interacting open strings becomes more
singular with increasing $p$, and for $p\geq 5$ the entropy diverges at
the transition. This implies that interaction effects become
important already for $T< T_H$. Nevertheless, it is likely that
the high temperature phase again involves liberated long strings.
We may argue for this as follows. The free energy of non-interacting
open strings is of order $G_o^0$, and interactions are unlikely
to change this scaling. On the other hand, the free energy
of liberated strings, (\ref{exten}), is of order $G_o^{-2}$.
Therefore, for weak coupling and for $T$ sufficiently above $T_H$,
the system can lower its free energy by liberating long
strings from the bound state. It is not clear, however, whether
the transition for $p>4$ is second order; it may be
a first order transition for all values of $G_o$. Additional
ideas on the Hagedorn transition for 5-branes have appeared in
\cite{Berkooz,HarmarkZero,HarmarkOne}.

In fact,
one may suspect that for large enough $p$ the D$p$-brane does not decouple
from gravity in the NCOS limit. In \cite{GMMS} the non-planar one-loop
amplitude was calculated for 4 open strings, 
and it was shown that for $p< 7$ the amplitude is finite for
$k^2=0$ ($k$ is the momentum in the closed string channel). 
For $p \geq 7$ the amplitude blows up for $k^2=0$
which suggests that there is no decoupling from massless
bulk modes. In order to check this, we have
calculated the cross-section for two massless open strings 
of energy $k_0$ colliding
along the electric field direction to produce an outgoing graviton.
The term in the Born-Infeld action describing this process is
$$ {1\over 2}\int d^{p+1} x\ 
(\partial_0 \Phi^i \partial_0 \Phi^j- 
\partial_1 \Phi^i \partial_1 \Phi^j)
(\delta_{ij} + \sqrt 2 \kappa h_{ij})
\ ,
$$
where we have rescaled the scalar fields so that they are canonically 
normalized. The cross-section we find,
$$\sigma \sim G_o^4 (\alpha'_e)^4  k_0^{9-p}
(E_c^2 - E^2)^{(7-p)/2}\ ,
$$
vanishes for $p<7$, is finite for $p=7$, and diverges for $p>7$.
This result is consistent with the annulus calculation in \cite{GMMS}
and it indicates that non-gravitational NCOS theories can
exist only for $p<7$.
It is interesting to note that $p=7$ is also special from the point of view
of the thermodynamics: indeed here the free energy for the
low energy phase, computed in section~\ref{LowT}, diverges
logarithmically at $T=T_H$.

To summarize this section, we can draw the general 
conclusion that the physics of the
Hagedorn transition is similar for all NCOS theories with $p<7$,
in that above $T_H$ the temperature dependence
of $F$ is effectively
two-dimensional even though $F$ is extensive in $p$ dimensions.  This is
similar to the answer proposed in {\it cf.} \cite{AW}, although the
justification there was different.
In the NCOS case the two-dimensional behavior of the free
energy above the Hagedorn temperature has to do with the presence of
the electric field.

\subsection{Extension to OD3 theory}
\label{OM}

In NCOS theory, a critical NS-NS 2-form field in presence of a
D$p$-brane leads to a decoupling limit in which fundamental strings
are light.  In certain variants of OM theory, a critical RR
$(p+1)$-form field applied to an NS5-brane is associated with a
decoupling limit in which D$p$-branes become light
\cite{GMSS,HarmarkTwo}.  No computational framework comparable to
perturbative quantization of strings has emerged to study light
D$p$-branes for $p>1$.  Indeed, one may wonder if it is logically
consistent for higher dimensional branes ever to be the
``fundamental'' degrees of freedom of a theory.\footnote{It is
unquestionable that $p$-branes on shrinking cycles play a role in
gauge symmetry enhancement, as well as in elucidating singularities
like the conifold.  What seems less certain is whether some theory in
non-compact spacetime exists which admits a fundamental description as
a theory of fluctuating $p$-branes, $p>1$.}  However, it appears from
the decoupling arguments of \cite{GMSS,HarmarkTwo} that there are
decoupling limits of string theory where the lightest excitations are
indeed open D$p$-branes with $p>1$.

It is tempting to adapt the reasoning used for NCOS theories to
describe a possible phase transition for various OM-theories.  In this
section we will make an attempt in this direction, but our arguments
will be much more heuristic than in previous sections.  We will
examine the relatively clean example of OD3-theory, which is the
theory of open D3-branes on an NS5-brane in a decoupling limit with a
critical 4-form potential turned on.  Besides the obvious
pitfall that the quantum states of fluctuating open D3-branes are hard
to count, there is another interesting effect: fundamental strings
living on the NS5-D3 bound state have a substantially reduced tension
relative to their tension in flat space, 
and their Hagedorn behavior competes with the tendency to liberate
D3-branes.

An NS5-brane with a near-critical RR 4-form potential can be
described as a bound state of many D3-branes and a single NS5, such
that the D3-branes make the dominant contribution to the tension.  Let
$\rho$ be the number density of D3-branes in the two directions
orthogonal to the D3-branes but parallel to the NS5-brane.  Then we
require $\rho \tau_{D3} \gg \tau_{NS5}$.  We will see below that this
condition turns out to be trivially satisfied in the OD3 limit as
defined in \cite{GMSS}.

The tension of the NS5-D3 bound state is $\sqrt{\tau_{NS5}^2 + \rho^2
\tau_{D3}^2}$.  The tension of an open D3-brane stuck to the NS5-brane
is
  \eqn{TOD}{\eqalign{
   \tau_{OD3} &= {d \over d (\delta\rho)} \left( \delta\rho \tau_{D3} + 
      \sqrt{\tau_{NS5}^2 + (\rho-\delta\rho)^2 \tau_{D3}^2} \right)
      \Bigg|_{\delta\rho=0}  \cr
    &= \tau_{D3} \left[ 1 - \left( 1 + 
      \left( {\tau_{NS5} \over \rho \tau_{D3}} \right)^2 \right)^{-1/2}
     \right]  \cr
    &\approx {1 \over 2} \left( {\tau_{NS5} \over \rho \tau_{D3}} \right)^2
      \tau_{D3} \,,
  }}
 where in the last line we have used $\rho\tau_{D3} \gg \tau_{NS5}$.
The near-critical scaling limit is described by two parameters
\cite{GMSS}: a scale $\tilde\alpha'_{\rm eff}$ and a coupling
$G_{o(3)}^2$, which happens to be precisely the closed string coupling
$g_{\rm str}$ (this last fact is special to OD3-theory).  The precise
scaling of the parameters is given as: 
$\tilde\alpha' = \sqrt{\epsilon} \, \tilde\alpha'_{\rm eff}$, the
metric in directions transverse to the D3-branes scales as $g_{MN} =
\e \d_{MN}$, and $g_s = G_{o(3)}^2$.  The scaling of the metric implies
that $\rho = \frac{\rho_0}{\e}$, where $\rho_0$ is of order unity.
This implies $\rho \tau_{D3} \sim O\left( {\e^{-2}} \right)$, while
$\tau_{NS5} \sim O\left( {\e^{-\frac{3}{2}}}\right)$, thereby
satisfying the aforementioned condition that $\rho\tau_{D3} \gg
\tau_{NS5}$.

Just as we found in NCOS theory that it is thermodynamically favorable
to liberate strings from the bound state at a temperature $T_H \sim
\sqrt{\tau_{\rm eff}}$, so we will find here that it is favorable to
liberate D3-branes at a temperature $T_{c,D3} \sim \tau_{OD3}^{1/4}$.
The argument proceeds along similar lines.  First we note that a free
$U(1)$ gauge multiplet in a flat-space theory in $p+1$ dimensions and
sixteen supercharges has
  \eqn{FreeAnyDim}{
   {F \over L^p T^{p+1}} = -c_{Dp} \equiv
    -{8 {\rm Vol} \, S^{p-1} \over (2\pi)^p}
        \left( 2 - {1 \over 2^p} \right) \Gamma(p) \zeta(p+1) \,.
  }
 Here $F$ is the free energy, $L^p$ is the spatial world-volume, and
$T$ is the temperature.  The low-energy dynamics of the bound state is
non-commutative super-Yang-Mills theory in 5+1 dimensions.  Clearly,
then, the free energy at low temperatures is order $1$ in the sense
that it does not grow with a power of the number density $\rho$.  Let
us assume that this remains the case up through the temperature where
liberating D3-branes becomes thermodynamically favorable.  Then the
same manipulations that we went through in \FkForm\ are justified at
large $\rho$: the free energy after a number density $\delta\rho$ of
D3-branes have been liberated is
  \eqn{FrhoForm}{\eqalign{
   {F_{\delta\rho} \over L^5} &= \sqrt{\tau_{NS5}^2 +
     (\rho-\delta\rho)^2 \tau_{D3}^2} + \delta\rho \, \tau_{D3} -
     c_{D3} \delta\rho \, T^4  \cr
    &\approx \rho \tau_{D3} - c_{D3} \rho T^4 +
      {\tau_{NS5}^2 \over 2 (\rho-\delta\rho) \tau_{D3}} +
      c_{D3} (\rho-\delta\rho) T^4  \cr
    &\geq \rho \tau_{D3} + \tau_{NS5} \sqrt{2 c_{D3} T^4/\tau_{D3}} -
       c_{D3} \rho T^4 \,,
  }}
 where we neglect terms which are subleading in $\rho$.  Equality
pertains in the last line of \FrhoForm\ if and only if the last two
terms in the second line are equal.  One reads off the fraction of
liberated D3-branes, the free energy, and the critical temperature
$T_{c,D3}$ as
  \eqn{AllInfo}{\eqalign{
   \nu \equiv {\delta\rho \over \rho} &= {T^2 - T_{c,D3}^2 \over T^2}  \cr
   {F \over L^5} &\approx \rho \tau_{D3} - c_{D3} \rho (T^2 - T_{c,D3}^2)^2
     \cr
   T_{c,D3}^2 &= {\tau_{NS5} \over \sqrt{2 c_{D3} \tau_{D3}} \rho } 
               = \sqrt{\tau_{OD3}/c_{D3}} 
               = \frac{1}{(2 \pi)^{\frac{7}{2}}} \, \frac{1}{\sqrt{2 c_{D3}}}
		\, \frac{1}{g_{str}^{3/2} \tilde{\a}'^2 \rho_0} \,.
  }}
 (Formally, a similar analysis seems to be possible for many D5-branes
bound to an NS5-brane.  However, in this case, the absence of strong
IR dynamics on $N$ coincident D5-branes makes it likely that there are
$O(N^2)$ massless degrees of freedom even at low energies.  This would
overwhelm the $O(N)$ effect due to liberated D5-branes, rendering the
whole approach suspect.)

In the case of NCOS theory, it was essentially guaranteed that
fundamental strings would start being liberated at the Hagedorn
temperature of the non-commutative open strings, because the
calculation of the free energy of liberated strings was equivalent to
a computation in the light cone formalism of highly excited open
strings.  In OD3 theory, no analogue of the latter computation exists
as yet, so to be conservative we should regard $T_{c,D3}$ as an upper
bound on the temperature where some transition must take place.  In
fact, as we will now show, when $G_{o(3)} \ll 1$, there is a Hagedorn
transition for closed fundamental strings living on the NS5-D3 system
at a substantially lower temperature than $T_{c,D3}$.  These closed
strings are excitations of the NS5-D3 bound state.  

There is no net f1 charge in the NS5-D3 bound state that we wish to
analyze; however, in order to extract the tension of the closed
strings which live on the NS5-D3 system, it is convenient to first
consider a BPS arrangement where an NS5-brane is oriented in the
012345 directions, $\rho$ D3-branes per unit 45-volume are oriented in
the 0123 directions, and $\rho_4$ fundamental strings per unit
2345-volume are oriented in the 01 directions.  The total tension is
  \eqn{TotalTau}{
   \tau = \sqrt{(\tau_{NS5} + \rho_4\tau_{f1})^2 + (\rho\tau_{D3})^2} \,.
  }
 The effective tension of a fundamental string bound to the NS5-D3
system is
  \eqn{EffectiveTauF}{
   \tau_{f1,\rm eff} = \left. {\partial\tau \over \partial\rho_4}
    \right|_{\rho_4=0} = \tau_{f1} \left. {\tau_{NS5} \over \tau}
    \right|_{\rho_4=0} \approx
     \tau_{f1} {\tau_{NS5} \over \rho \tau_{D3}} \ll \tau_{f1} \,,
  }
 where in the last two steps we have used the fact that most of the
mass of the bound state is carried by the D3-branes.  Fundamental
strings which are orthogonal to the D3-branes but contained in the
NS5-branes are much heavier.

Another way to derive the effective tension $\tau_{f1,\rm eff}$ is to
look at the supergravity solution for the NS5-D3 system.  The string
metric and dilaton are 
  \eqn{DNSsoln}{\eqalign{
   ds_{\rm str}^2 &= {1 \over \sqrt{h_3}} (-dt^2 + dx_1^2 + dx_2^2 + dx_3^2)
    \cr &\qquad{} + \sqrt{h_3} \left( dx_4^2 + dx_5^2 + 
      h_5 (dx_6^2 + dx_7^2 + dx_8^2 + dx_9^2) \right)  \cr
    e^{2(\phi-\phi_\infty)} &= h_5 \qquad
    h_3 h_5 = 1 + {q_3 \over r^2} \qquad h_5 = 1 + {q_5 \over r^2}  \cr
    r^2 &= x_6^2 + x_7^2 + x_8^2 + x_9^2 \,.
  }}
 In the limit where $\rho\tau_{D3} \gg \tau_{NS5}$, we have $q_5/q_3 =
\tau_{NS5}/(\rho \tau_{D3})$.  The three-form field strengths need not
concern us, except to note that $B_{\mu\nu}^{(NS)}$ may be chosen so
that only $B_{45}$ is non-zero.  We are considering many D3-branes,
but only a single NS5-brane, so the supergravity solution is
trustworthy, in the sense that curvatures are sub-stringy, down to a
radius $r_{\rm match} = \sqrt{q_5}$.  Following the philosophy of
\cite{hp}, we assert that the tension and coupling of fundamental
strings bound to the NS5-D3 system can be read off, up to factors of
order unity, from the properties of a test string located at the
matching radius $r_{\rm match}$.  The tension so derived agrees with
\EffectiveTauF.  The advantage of this more heuristic approach is that
we can extract the string coupling for the strings bound to the NS5-D3
system: up to a factor of $2$ it is just $g_{\rm str} = G_{o(3)}^2$.

When $G_{o(3)} \ll 1$, we are entitled to use the free string spectrum
to predict a Hagedorn temperature for the light fundamental strings
whose orientation is within the D3-branes.  It is
  \eqn{TcfDef}{
   T_{c,f1} \sim \sqrt{\tau_{f1,\rm eff}} = \frac{1}{(2 \pi)^{\frac{3}{2}}}
    \frac{1}{\sqrt{g_{str} \tilde{\a}'^2 \rho_0}} \,.
  }
 From 
  \eqn{TRatio}{
   {T_{c,f1} \over T_{c,D3}} \sim {\sqrt{\tau_{f1, \rm eff}} \over
    \tau_{OD3}^{1/4}} \sim g_{\rm str}^{1/4} = \sqrt{G_{o(3)}} \,,
  }
 we learn that the fundamental string Hagedorn transition
happens at a lower temperature when $G_{o(3)} \ll 1$.  This supports
the view that the most relevant degrees of freedom in 
OD3-theory with $G_{o(3)} \ll 1$ may be little strings.  Before one can ask
whether the D3-brane liberation transitions occurs, one must
understand what contribution the fundamental strings make to the free
energy above $T_{c,f1}$.

When $G_{o(3)} \gg 1$, one can obtain a more natural description of
the theory by S-dualizing.  The scaling limit leading to OD3-theory
S-dualizes into the zero slope limit used in \cite{SW} to obtain
non-commutative Yang-Mills theory \cite{GMSS}.  For $G_{o(3)} \gg 1$,
this theory is weakly coupled in the sense that $g_{YM} \ll
\sqrt{\theta}$.  However, the interacting theory is
non-renormalizable, so it might be inappropriate to regard the quanta of
the gauge field as the fundamental degrees of freedom.  It was
suggested in \cite{GMSS} that OD3-theory provides an ultraviolet
completion of 5+1-dimensional non-commutative Yang-Mills theory.  This
is not a very effective description in the absence of a knowledge of
how to quantize open
D3-branes.  It must be admitted that, for $G_{o(3)} \gsim
1$, there is neither a natural little string theory description of
OD3-theory, nor a renormalizable interacting quantum field theory
description.  Despite all this uncertainty, the analysis following
\FrhoForm\ may still be valid: it only depends on the free energy of
the bound state being $O(\rho^0)$ in a $\rho \to \infty$ limit with
$G_{o(3)}$ fixed.

\section*{Acknowledgements}

We are grateful to J.~Maldacena, E.~Rabinovici and S.~Shenker
for useful discussions.  The work of
S.S.G.\ was supported in part by DOE grant~DE-FG02-91ER40671, and by a
DOE Outstanding Junior Investigator award.
The work of S.G. was supported in part by the Caltech Discovery Fund,
grant RFBR No 98-02-16575 and Russian President's grant No 96-15-96939.
The work of I.R.K.\ was
supported in part by the NSF grant PHY-9802484 and by the James
S.~McDonnell Foundation Grant No. 91-48. M.R.\ was supported in part by
NSF grant PHY-980248 and by the Caltech Discovery Fund.
S.S.G.\ and I.R.K.\ thank
the Aspen Center for Physics for hospitality while this work was in
progress.

\begingroup\raggedright\endgroup

\end{document}